\documentclass[12pt,preprint]{aastex}
\def \etal{{\it et al.}}
\def \eg{{\it e.g.,}}

\def \TRACE{{\it TRACE}}
\def \SOHO{{\it SOHO}}
\def \Yohkoh{{\it Yohkoh}}

\shorttitle{CDS Coronal Loops}
\shortauthors{J.T. Schmelz et al.}

\begin{document}

\title{All Coronal Loops are the Same: Evidence to the Contrary}

\author{J.T. Schmelz, K. Nasraoui, V.L. Richardson, P.J. Hubbard, C.R. Nevels, J.E. Beene}
\affil{Physics Department, University of Memphis, Memphis, TN 38152}
\email{jschmelz@memphis.edu }

\begin{abstract}

The 1998 April 20 spectral line data from the Coronal Diagnostics Spectrometer (CDS) on the {\it Solar and Heliospheric Observatory} (\SOHO) shows a coronal loop on the solar limb. Our original analysis of these data showed that the plasma was multi-thermal, both along the length of the loop and along the line of sight. However, more recent results by other authors indicate that background subtraction might change these conclusions, so we consider the effect of background subtraction on our analysis. We show Emission Measure (EM) Loci plots of three representative pixels: loop apex, upper leg, and lower leg. Comparisons of the original and background-subtracted intensities show that the EM Loci are more tightly clustered after background subtraction, but that the plasma is still not well represented by an isothermal model. Our results taken together with those of other authors indicate that a variety of temperature structures may be present within loops.

\end{abstract}


\section{Introduction}

The temperature distribution along coronal loops provides a basic observable to be predicted by any coronal heating model, so measuring this distribution using data from the current fleet of solar imagers and spectrometers has become a priority. Unfortunately, new results have not led to a consistent picture. Analysis of coronal data obtained with the EUV Imaging Telescope (EIT) on \SOHO\ and the Transition Region and Coronal Explorer (\TRACE) seems to imply that loops have a uniform temperature (\eg\ Neupert \etal\ 1998; Lenz \etal\ 1999). Analysis of the data from the Soft X-ray Telescope (SXT) on \Yohkoh\ and \SOHO -CDS shows, however, that loop temperatures increase from the footpoints to the apex (\eg\ Kano \& Tsuneta 1996; Schmelz \etal\ 2001 -- hereafter Paper I).

More recent results have established the importance of background subtraction for coronal loop analysis -- separating the actual loop plasma from the diffuse foreground/background emission that results from unresolved coronal structures and/or instrumental effects. Background-subtracted CDS loop results (Del Zanna \& Mason 2003; DiGiorgio, Reale \& Peres 2003; Landi \& Landini 2004) are sometimes (but not always) consistent with isothermal plasma along the line of sight and/or along the length of the loop. The results described in Paper I implied multithermal plasma in both directions but did not include background subtraction. In this paper, therefore, we add this step to our original analysis and describe the resulting similarities to and differences from previously published CDS results.

\section{Observations and Analysis}

The data used here were taken with CDS (Harrison \etal\ 1995) on 1998 April 20 at 20:54 UT and are described in detail in Paper I. Pauluhn \etal\ (1999) determined the point spread function of CDS ($6''\times 8''$), and the loop alignments discussed below are within this tolerance. There was a coronal mass ejection at 10:07 UT on 1998 April 20, but movies of the loop formation and evolution made in the EIT 195 \AA\ passband and the SXT AlMg filter show that the loop was stable during the CDS observations. Unfortunately, \TRACE\ was not yet taking scientific data, and EIT and SXT were each imaging in only one filter, so no temperature diagnostics were available from these instruments. It was possible, however, to use the closest SXT image as a high-temperature constraint for the Differential Emission Measure (DEM) analysis, as was described in Paper I. 

Figure 1 shows multiple wavelength frames for our CDS loop. The loop is seen most clearly in the hotter temperature lines (Mg IX to Fe XVI). There is also a companion loop seen slightly to the south of the main loop (see Fe XII window) that passes either in front of or behind the main loop. We avoided this position in our analysis. There is certainly cool plasma (see O V, Ne VI, and Ca X frames) above the limb, and the analysis described below will determine if it is part of the main loop structure. We examined all available wavelength images in detail to choose pixels where there appeared to be little background contamination. We chose three pixels along the visual center of the loop as well as a pair of background pixels for each loop pixel, one inside the loop and one outside. These positions are shown in Figure 2 against the Si XII image. 

Our background subtraction results are shown in Table 1 for the set of pixels near the loop apex. The first four columns list a running line number, the ion, the wavelength in \AA, and the log of the peak formation temperature. The next three columns list the intensities of the loop apex pixel, the outside background pixel, and the inside background pixel. The intensities and associated uncertainties were obtained with the CDS analysis program FITLINE available in {\t SolarSoft} and are in units of ergs cm$^{-2}$ s$^{-1}$ sr$^{-1}$. Column eight lists the background-subtracted intensity, which is obtained by subtracting the average of the inside and outside background pixel intensities from the corresponding loop pixel intensity. The propagated errors are listed in terms of $\sigma$ in the last column of the table. Resulting intensities above 5$\sigma$ were taken to be significant and are used in the background-subtracted plots described below. The fact that some of these spectral lines did not have significant intensities after this subtraction process may indicate that they were part of the unresolved background emission rather than part of the target loop.

We used the atomic physics calculations compiled in version 4.02 of the CHIANTI Atomic Physics database (Dere \etal\ 1997; Young \etal\ 2003), the ionization fractions of Arnaud \& Raymond (1992) for the iron lines and Mazzotta \etal\ (1997) for all other lines, and the ``hybrid'' elemental abundances from Fludra \& Schmelz (1999). Densities measured from density-sensitive line ratios gave average values of 1e9, 2e9, and 2e9 cm$^{-3}$ for the apex, upper-leg, and lower-leg pixels, respectively. These values were used to create the EM Loci plots (Jordan \etal\ 1987) shown in Figure 3. The left-hand column shows the original results (before background subtraction) where each curve represents the possible temperature-emission measure solutions for one of the spectral lines listed in Table 1. Since these curves do not intersect at a single point, the plasma along the line of sight cannot be isothermal. The plots in the right-hand column show the results after background subtraction. There are fewer curves -- only the 5$\sigma$ results from Table 1 are used. The curves certainly cluster more tightly here than in the original plots on the left. We also adjusted the elemental abundances of the non-iron lines to obtain the tightest possible clustering, but the curves still did not intersect at a single point. Therefore, we must conclude that the background-subtracted plasma is not isothermal. Possible explanations for the observed multithermal plasma include a series of isothermal loops contributing along the line of sight, or multiple adjacent isothermal strands at different temperatures within the resolution element. We defer the full DEM treatment of these background-subtracted data to a future paper.

\section{Discussion}

The background-subtracted intensities for our loop show that the loop apex is visible in a variety of coronal lines, including Fe XII, Fe XIII, Fe XIV, and Fe XVI. The upper-leg pixel shows these lines as well as several slightly cooler coronal lines. For the lower-leg pixel, the cooler transition region O V line is also visible, but the hottest (Fe XVI) line is not. These data suggest that the overall temperature increases from the footpoints to the loop top. So, although background subtraction has affected the details of the analysis, two of the original results from Paper I still stand -- the plasma is multi-thermal both along the length of the loop as well as along the line of sight. Since these results disagree with some of the CDS loop results published recently, it is important to look at these differences in more detail. 

The loop leg studied by Del Zanna \& Mason (2003) is clearly visible in the upper transition region CDS lines such as Mg VII, Ca X, and Mg IX as well as the \TRACE\ 171-\AA\ passband. The structure in not visible in higher-temperature CDS lines such as Fe XII and Fe XIII or in cooler transition region lines like O V. These CDS data were consistent with cooler ($T = 0.7-1.1 \times\ 10^6$ K), isothermal plasma along most lines of sight investigated. This result is clearly different than ours. On the other hand, these authors also found that the loop had steeper temperature and density gradients along its length than those derived from previous studies based on \TRACE\ observations alone. This last result is similar to ours.

The hotter loop investigated by DiGiorgio, Reale \& Peres (2003) has some similarities to the one studied here. The bulk of their loop was visible in the hotter CDS lines (Fe XII, Fe XIV, and Fe XVI), but contrary to our result, only the footpoints of their loop were visible in the intermediate temperature CDS lines (Mg IX and Mg X). An EM Loci plot of their hottest lines in the region of the loop apex yielded a temperature of about 2 MK, again similar to ours. Filter ratios of their loop from \Yohkoh\ images gave a temperature closer to 4 MK, however. This hotter plasma component could indicate a multithermal structure along the line of sight, but given the time interval between the \Yohkoh\ and CDS data, the possibility could not be excluded that CDS was observing plasma that was slowly cooling. 

Landi \& Landini (2004) analyzed a loop most clearly visible in a pair of Fe XVI CDS lines. The loop is also visible in slightly cooler lines (Fe XII, Fe XIII, Fe XIV), but only the footpoints are visible in even cooler lines (Mg VIII and Mg IX line). This characteristic makes this loop similar to the loop observed by DiGiorgio, Reale \& Peres (2003), but different from our loop. Landi \& Landini (2004) used EM Loci plots to show that the plasma is isothermal along most of the lines of sight investigated, but not at the loop footpoints. We note that, when the Landi \& Landini loop and our loop are scaled to have the same length, our DEM-weighted temperatures (Paper I) actually fit inside the error bars in their Figure 8, which shows the loop temperature as a function of arc length (Landi 2002, private communication).

These CDS results all assume that the CHIANTI data are correct. If we were to adjust these data so the EM Loci plots of Figure 3 were consistent with an isothermal plasma, then we must also change the isothermal results like those described in the last few paragraphs. For example, the atomic data for our Fe~XVI line at 360 \AA\ would have to be adjusted, but if this same change were made to the EM Loci plots of DiGiorgio, Reale \& Peres (2003) or Landi \& Landini (2004), their results would not be ``as isothermal'' as before. A careful search through the CDS atlas reveals multiple examples of the various types of loops described above. These results indicate that CDS loops have different temperature characteristics and, perhaps, that different heating mechanisms may be operating to produce them.

We would like to thank Jonathan Cirtain, Julia Saba, and Amy Winebarger. Support was provided by NASA grants NAG5-9783 and NAG5-12096. \SOHO\ is a project of international cooperation between ESA and NASA. CHIANTI is a collaborative project involving the NRL (USA), RAL (UK), and the Universities of Florence (Italy) and Cambridge (UK).

{}

\begin{figure}
\figurenum{1}
\epsscale{.9}
\plotone{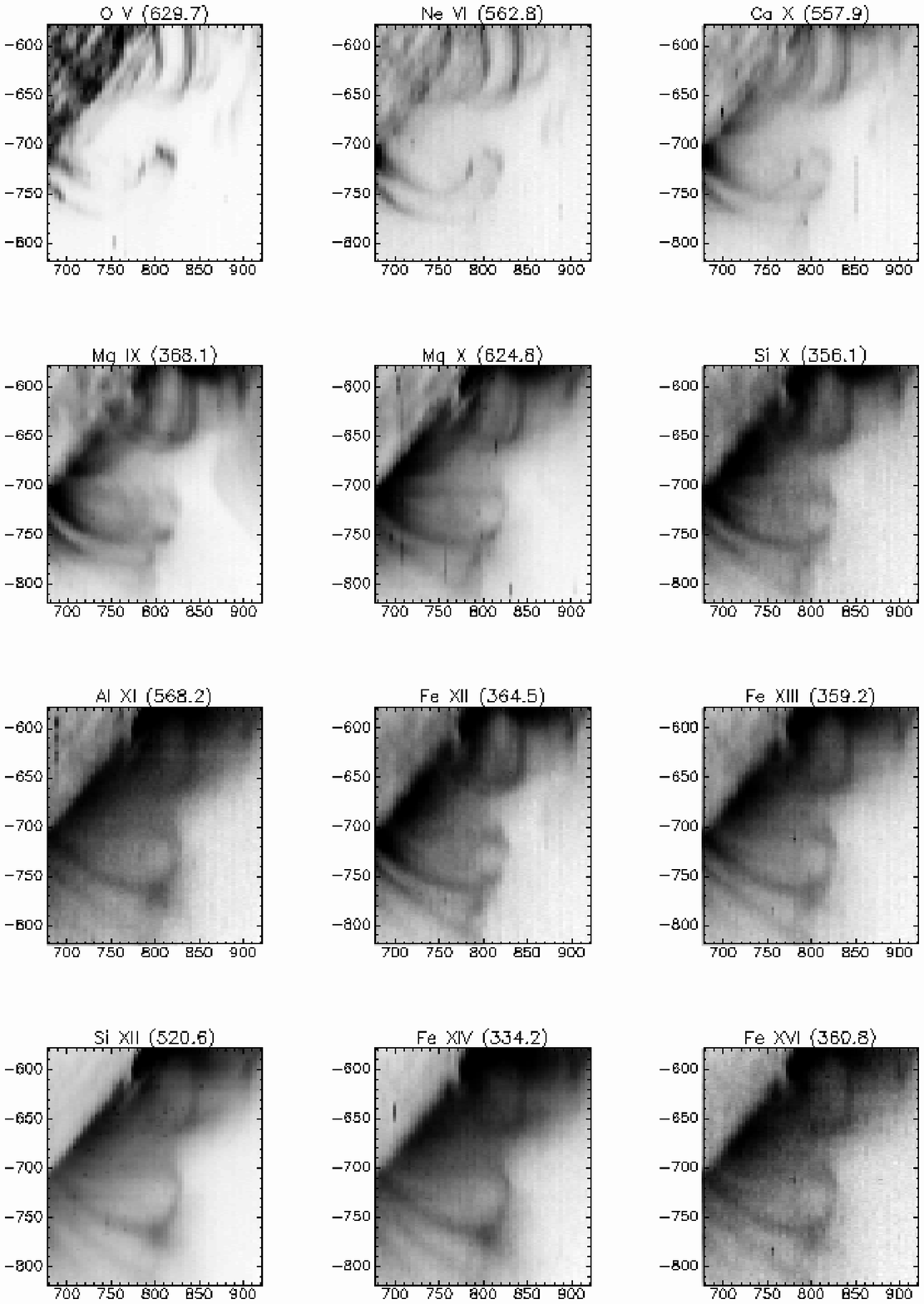}
%\plottwo{<epsfile>}{<epsfile>}
\caption{Co-aligned CDS loop data from 1998 April 20 at 20:54 UT seen here on the southwest limb at solar coordinates (790$''$, -690$''$). All frames are labeled with the appropriate ion and its corresponding wavelength, and are arranged in order of peak formation temperature, from coolest to hottest (see Table 1 for details). The images depict the monochromatic peak intensity of the line listed (not the summed intensity over the entire CDS wavelength window). The intensity scale has been inverted so the loops appear dark in these images. Cosmic ray hits were flagged as missing data and were not included in the analysis.}
\end{figure}

\begin{figure}
\figurenum{2}
%\epsscale{.8}
\includegraphics[angle=90]{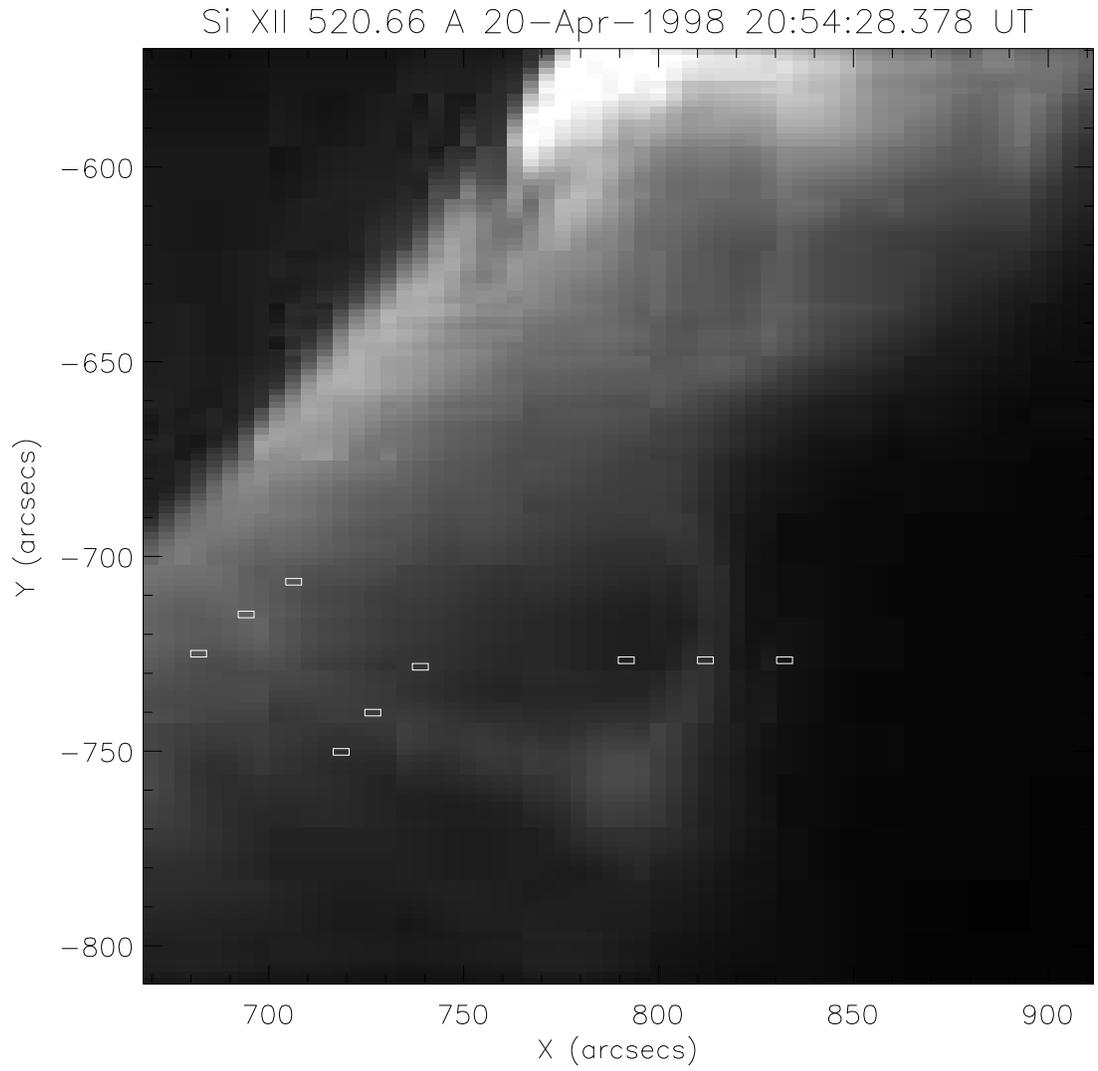}
%\plotone{f2.eps}
%\plottwo{<epsfile>}{<epsfile>}
\caption{Close-up of the CDS Si XII window (the bottom left panel of Figure 1, but with a positive intensity scale). Si XII has a peak formation temperature of Log T $=$ 6.25. The pixels examined in detail are marked with open white rectangles.}
\end{figure}

\begin{figure}
\figurenum{3}
\epsscale{.8}
\plotone{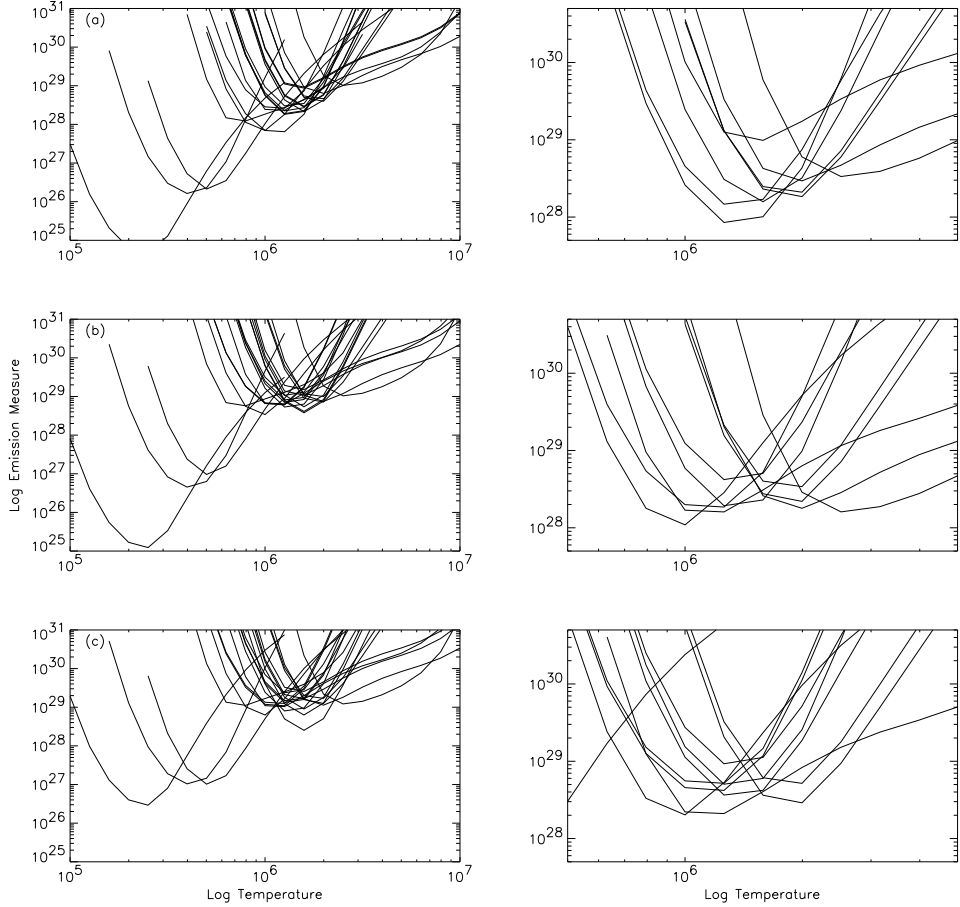}
%\plottwo{<epsfile>}{<epsfile>}
\caption{ EM Loci plots for the (a) apex pixel, (b) upper-leg pixel, and (c) lower-leg pixel. The plots displayed in the left-hand column show curves using the original intensities (before background subtraction) for all spectral lines listed in Table 1. The plots in the right-hand column show curves for the spectral lines with significant intensity ($>$ 5 sigma in Table 1) after background subtraction. Note the different scales used in the left- and right-hand plots. Each x-axis tick mark represents an increment of Log T $=$ 0.1.}
\end{figure}

\begin{deluxetable}{lllllllll}
\tabletypesize{\scriptsize}
\tablewidth{0pt}
\tablehead{
\colhead{} & \colhead{Ion} & \colhead{$\lambda$} & \colhead{Log T} & 
\colhead{Loop$^1$} & \colhead{Outside$^1$ } & \colhead{Inside$^1$ } &
\colhead{Subtracted$^1$} & \colhead{Sigma$^2$} 
}
\startdata

1&O V &629.73 &5.40 &5.18$\pm$0.92 &---- &22.3$\pm$2.04 &-5.95$\pm$2.24 &----\\
2&Ne VI&562.80&5.65&5.11$\pm$1.19&----&9.90$\pm$1.90&0.16$\pm$2.24 &0.07\\
3&Ne VII&561.73&5.70&1.02$\pm$0.73&----&5.11$\pm$1.70&-1.53$\pm$1.85 &----\\
4&Ca X&557.77&5.80&10.2$\pm$1.42&----&12.6$\pm$1.62&3.92$\pm$2.15 &1.82\\
5&Mg IX&368.07&6.00&207.$\pm$4.73&60.3$\pm$2.57&293.$\pm$5.53& 31.1$\pm$7.72 &4.03\\
6&Mg X&624.94&6.05&95.1$\pm$3.60&42.1$\pm$2.51&91.3$\pm$3.69& 28.4$\pm$5.73 &4.95\\
7&Si IX&341.95&6.05&28.2$\pm$2.22&28.3$\pm$2.29&28.7$\pm$2.34&-0.33$\pm$3.96&----\\
8&Si IX&349.87&6.05&20.1$\pm$2.06&18.6$\pm$2.00&55.9$\pm$2.80&-17.1$\pm$4.01 &----\\
9&Si X&356.01&6.10&46.3$\pm$2.67&----&63.9$\pm$3.06&14.4$\pm$4.06 &3.54\\
10&Al XI&550.03&6.15&25.8$\pm$2.13&9.58$\pm$1.24&17.7$\pm$1.79&12.1$\pm$3.05 &3.99\\
11&Al XI&568.12&6.15&56.1$\pm$2.88&21.3$\pm$1.96&32.4$\pm$2.36&29.2$\pm$4.21 &6.94\\
12&Fe XII&346.85&6.15&24.2$\pm$2.03&7.72$\pm$1.64&21.6$\pm$2.55&9.55$\pm$3.65 &2.62\\
13&Fe XII&364.46&6.15&65.2$\pm$3.06&11.7$\pm$1.80&57.7$\pm$3.06&30.47$\pm$4.69 &6.50\\
14&Si X&347.40&6.10&77.6$\pm$3.17&22.1$\pm$1.96&57.7$\pm$3.36&37.7$\pm$5.02 &7.52\\
15&Fe XIII&320.81&6.20&35.3$\pm$2.66&12.4$\pm$2.11&18.8$\pm$5.47&19.7$\pm$6.44 &3.05\\
16&Fe XIII&321.40&6.20&21.6$\pm$2.84&10.8$\pm$1.77&15.4$\pm$3.16&8.49$\pm$4.60 &1.84\\
17&Fe XIII&348.18&6.20&78.7$\pm$3.31&20.1$\pm$2.02&49.5$\pm$3.08&44.0$\pm$4.95 &8.88\\
18&Fe XIV&334.17&6.25&181.$\pm$5.12&61.7$\pm$3.16&133.$\pm$4.60&83.8$\pm$7.57 &11.1\\
19&Fe XIV&353.83&6.25&60.4$\pm$3.13&20.1$\pm$1.63&45.7$\pm$2.66&27.5$\pm$4.42 &6.22\\
20&Si XII&520.67&6.25&191.6$\pm$5.18&80.4$\pm$3.33&125.$\pm$4.22&88.7$\pm$7.47 &11.9\\
21&Fe XVI &360.76&6.40&906.$\pm$9.35&534.$\pm$7.54&685.$\pm$7.88&296.$\pm$14.4 &20.6\\

\enddata

\tablenotetext{1}{Intensities and uncertainties in ergs cm$^{-2}$ s$^{-1}$ sr$^{-1}$}

\tablenotetext{2}{Significance ($\sigma$) of the result after background subtraction}
\end{deluxetable}

\end{document}